# Sistematic Mapping Protocol

## ESTIMATION ACCURACY ON SOFTWARE DEVELOPMENT USING AGILE TECHNOLOGIES

Marcelo Fransoy[a] y Fernando Pinciroli[b].

[a]Maestría en Ingeniería en Sistemas de Información, Facultad Regional Buenos Aires, Universidad Tecnológica Nacional, Argentina.
[b]Instituto de Investigaciones, Universidad Nacional de San Juan, Argentina.

# Contents

# 1. Introduction

There are techniques and tools to estimate software project effort, such as function point estimation, use case point estimation, COCOMO and COCOMO II, comparison estimation, PMI-PRM, among others. They are based on the estimation of different variables [1] that, in the case of fuzzy techniques, they are linguistic quantifiers for many or most of their variables [2][3][4]. In many cases the experts' opinion is essential [5]. Regarding risk analysis, the experts' contribution is crucial, especially on the risk hierarchy definition, and the probabilities of occurrence.

Just the fourth part of software projects succeed in terms of completion according to what is planned, estimated, and specified [6]. At the same time, IT projects nature creates a great number of risks [7]. A great percentage of projects that do not reach success is due to incorrect effort estimation of and wrong or none risk categorization [8].

Our interest on this work is to identify and categorize the estimation techniques and tools employed on software development projects using agile methodologies. On this matter, the main goal is to represent a state of the application of these techniques and tools emphasizing on the existing evidence about its effort estimation accuracy.

Through Evidence-Based Software Engineering (EBSE), it is expected to transform the information need into a question that may be answered, establishing the evidence to answer that question, and therefore critically assessing evidence to determine its validity [9]. Kitchenham et al. state that EBSE has the intention "to provide the means by which current best evidence from research can be integrated with practical experience and human values in the decision-making process regarding the development and maintenance of software" [9]. The present document describes the planning phase of a Systematic Mapping Study (SMS), used to organize the research field findings, based on Petersen et al. [10] recommendations.

The rest of the article is organized as follows: the research method is described in Section 2, the strategy to mitigate threats to validity is presented in Section 3 and, finally, we offer our conclusions in Section 4.

# 2. Research method

## 2.1. Goal and research questions

This work has as a goal to identify and classify the estimations techniques used in software development agile methodologies based on the results found, and to compare their estimation accuracies against those obtained in traditional software development methodologies.

The research questions set (RQ) is presented in Table 1.

Table 1. Research questions

| RQ# | Research question | Description |
|---|---|---|
| RQ1 | What kind of methodologies or life cycle model are mentioned? | A list of agile methodologies used, such as Scrum, waterfall, Kanban, Scrumban, XP, etc. |
| RQ2 | What are the estimation techniques used in agile software development methodologies? | A list of estimation techniques used on agile methodologies, such as Scrum, Kanban, Scrumban, XP, etc. |
| RQ3 | What do techniques employed in agile methodologies estimate? | A list of the aspects that are effectively estimated, for instance: size, effort, cost, time, etc. |
| RQ4 | What kind of techniques they are? | A list of the kind of techniques that are used, judgment, count or calculation techniques, or for its use on the early or late stages of the software development life cycle. |
| RQ5 | What evidence is there about estimation accuracy on agile methodologies? | Identify the evidence on estimation accuracy in every type of methodology. |
| RQ6 | What differences on estimation accuracy report agile methodologies against traditional methodologies? | List of the differences that are reported between these two types of methodologies. |
| RQ7 | Which of the techniques found were reported to be used in the industry? | List of estimation techniques found to be used in real-world settings. |
| RQ8 | What were the usage results of the techniques found in the industry? | Available results about the use of these techniques in real-world settings. |

A set of publication questions (PQ) are also included, in order to characterize the bibliographic and demographic space (**¡Error! La autoreferencia al marcador no es válida.**).

Table 2. Publication questions

| PQ# | Publication question | Description |
|---|---|---|
| PQ1 | Where have the studies been published? | Know the studies distribution by the type of location: conferences, journals, or workshops. |

| PQ# | Publication question | Description |
|---|---|---|
| PQ2 | In which year were they published? | Number of publications per year. |
| PQ3 | What are the most active countries? | List of countries where they were published. |

## 2.2. Search strategy and study selection

The search strategy selected includes three approaches to look for primary studies. The first one is a manual search on Google Scholar of literature systematic revisions about software estimations on agile methodologies. The second is an automatic search made through the on-line sources of scientific study (digital libraries and databases). Finally, the set of studies will be completed using the forward snowballing technique [11]. Figure 1 shows this strategy.

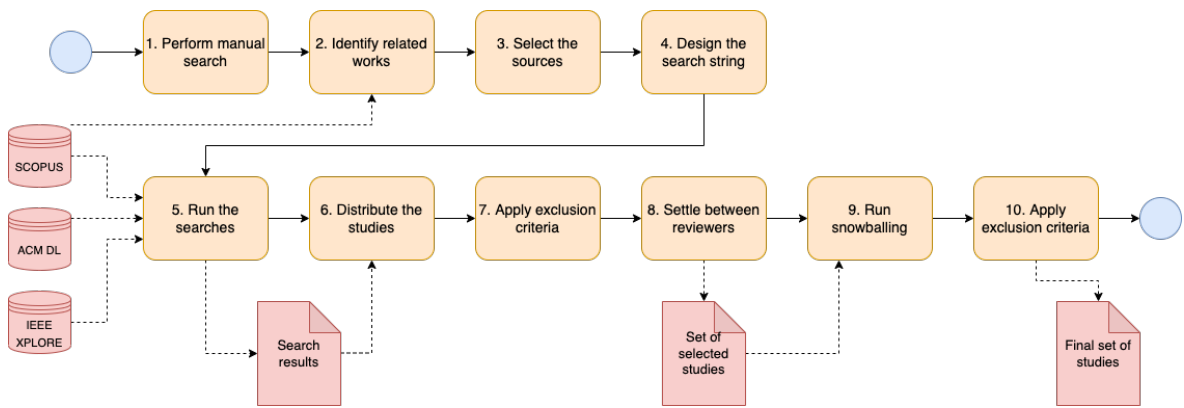

Figure 1. Search and selection process

The activities are as follows:

**Activity #1: Perform manual search**
A manual search on Google Scholar has been planned in order to obtain similar works to ours, according to Petersen et al. [12] recommendation, as it will also help to adjust the study approach.

**Activity #2: Identify related works**
The existing works about effort estimation in agile methodologies will be identified. Since we are performing a SMS, a secondary study, just other secondary works previously published will be considered as related works (SMS or SLR - Systematic Literature Review).

**Activity #3: Select the sources**
The electronic databases of scientific articles selected for this study are Scopus, IEEE Xplore and the ACM digital library, as they are repeatedly quoted at SMS reports and guidelines [13] [14] [15] [16].

**Activity #4: Design the search string for each source**
The search strings that will be used for the three mentioned libraries are indicated in Table 3.

Table 3. Search strings

| Database | Search string |
|---|---|
| Scopus | ( TITLE-ABS-KEY (agil* OR ágil OR scrum OR kanban OR scrumban OR xo OR "extreme programming") AND TITLE-ABS-KEY (estim* OR predic*)) AND PUBYEAR > 1999  AND ( LIMIT-TO(DOCTYPE,"ar" ) ) AND ( LIMIT-TO(SUBJAREA,"COMP " ) ) |
| IEEE Xplore | ("Index Terms":estim* OR "Index Terms":predic*) AND ("Index Terms":agil* OR "Index Terms":ágil OR "Index Terms":scrum OR "Index Terms":kanban OR "Index Terms":scrumban OR "Index Terms":xp OR "Index Terms":"extreme programming") |
| ACM | (Title:((agil* OR ágil OR scrum OR kanban OR scrumban OR xp OR "extreme programming")) AND Title:((estim* OR predic*))) OR (Keyword:((agil* OR ágil OR scrum OR kanban OR scrumban OR xp OR "extreme programming")) AND Keyword:((estim* OR predic*))) OR (Abstract:((agil* OR ágil OR scrum OR kanban OR scrumban OR xp OR "extreme programming")) AND Abstract:((estim* OR predic*))) |

**Activity #5: Run the searches**
The searches will be executed, and the results collected. These results will include duplicates that will have to be discarded by applying the following rules:
  a. Extended works: preserve the last one.
  b. Duplicated works: depending on the source, follow this order: Scopus (as it offers more detailed information), followed by IEEE Xplore and, lastly, ACM DL (because it does not recover the abstracts) [17].

**Activity #6: Distribute the studies**
The studies recovered will be distributed into four researchers as shown in table 4. Note that it is ensured that each work is examined by two different researchers in order to reduce biases.

Table 4. Distribution of studies.

| Researcher | Studies | | |
|---|---|---|---|
| | 0%-33% | 34%-67% | 68%-100% |
| R1 | X | | |
| R2 | | X | |
| R3 | | | X |
| R4 | X | X | X |

The individual selection of studies carried out by each researcher will be included into a unique set of studies.

**Activity #7: Apply exclusion criteria**
The researchers will independently revise the studies that have been assigned and they will decide if they are relevant or not, just by reading the title, abstract and by applying the exclusion criteria (EC). This criterion is described in table 6.

Table 5. Exclusion criteria.

| EC# | Description |
|---|---|
| EC1 | Short study (less than 5 pages). |
| EC2 | The study is not about software estimations. |
| EC3 | The study is not about agile software development. |
| EC4 | The study is not revised by peers. |

**Activity #8: Settle between reviewers**
The differences among researchers will be solved by using the following criteria [10]:

Table 6. Criteria to resolve disagreements.

| | | Researcher 1 | | |
|---|---|---|---|---|
| | | Include | Uncertain | Exclude |
| Researcher 2 | Include | A | B | D |
| | Uncertain | B | C | E |
| | Exclude | D | E | F |

**A & B:** the study is included.
**E & F:** the study is excluded.
**C & D:** the study is completely read and classified again till getting A, B, E or F.

**Activity #9: Run snowballing**
The resulting articles will be considered as "seed works" to be used on a forward snowballing technique, following the guidelines proposed by Wohlin [11]. The reason to execute this complementary search has as a goal to enrich the automatic search results.

**Activity #10: Apply exclusion criteria**
An expert will apply the exclusion criteria to the new resulting articles from the previous activity, therefore the definitive set of articles is obtained, which will be named as "selected articles" (SA).

## 2.3. Data extraction forms

The relevant data will be extracted from the set of studies in order to answer the eight RQ and the three PQ. The data will be stored on a spreadsheet with the format shown in table 7 and in table 8.

Table 7. Data extraction form for RQ.

| Study #ID | RQ1 | RQ2 | RQ3 | RQ4 | RQ5 | RQ6 | RQ7 | RQ8 |
|---|---|---|---|---|---|---|---|---|
| Study #1 | | | | | | | | |
| Study #2 | | | | | | | | |
| … | … | … | … | … | … | … | … | |
| Study #n | | | | | | | | |
| **Accepted values** | Methodology names | Techniques names | Aspects that are effectively estimated | Techniques names | Available results | Available results | Techniques name | Available results |

Table 8. Data extraction form for PQ.

| Study #ID | PQ1 | PQ2 | PQ3 |
|---|---|---|---|
| Study #1 | | | |
| Study #2 | | | |
| … | … | … | … |
| Study #n | | | |
| **Accepted values** | Conference Journal Workshop | Year of publication | Countries |

## 3. Threats to validity

Hereunder, the mitigation actions that have been expected are presented in order to reduce the impact of validity threats, categorized by Petersen [10], that could affect the study:

**Descriptive validity**

This validity aims to ensure that the observations are objectively and precisely described.

- The information to be collected has been organized by means of a pair of data extraction forms, for RQs and PQs, presented in Table 7 and Table 8 to support a uniform data register and objectify the data extraction process.
- Furthermore, all researchers will participate at an initial meeting, in order to standardize concepts and criteria, answer to any question and show (through examples) how to carry out the data extraction process.
- In addition, the data extraction form will also be published.

**Theoretical validity**

The theoretical validity depends on the capacity to obtain the information intended to get.
- It will start with a search string (Table 3) adapted to the three most popular digital libraries about computer science and software engineering online databases.
- An expert will provide a set of articles to verify if they are retrieved with the search string.
- A set of exclusion criteria has been defined (Table 6) to objectify the selection process.
- The studies will be distributed among four researchers, working independently and a study superposition ensuring that each of the studies is verified at least by two researchers (Table 4).
- Two different research methods will be combined: an automatic search and a manual search (snowball forward), to reduce the risk of not finding all the available evidence.

**Generalizability**

This validity is related to the capacity to generalize the results to the whole domain.
- The RQ set is sufficiently general to identify and classify the findings about estimation techniques on agile software development methodologies, independent of specific cases, the type of industry, etc. [12].
- The initial search of similar works will collaborate with the definition of generalized RQ.

**Interpretive validity**

This validity is achieved when the conclusions are reasonable considering the data.

- At least two researchers will validate each conclusion.
- Two researchers, with experience on the issue domain, will help with the data interpretation.

**Repeatability**

The research process must be detailed enough in order to ensure a thorough repetition.

- This protocol has been designed detailed enough as to allow the repetition of the process that has been followed.
- The protocol, as well as the study results, will be published online so that other researchers are able to replicate the process and confirm the results.

## 4. Conclusions

Regarding the SMS process, the guidelines published by Petersen [10] have been strictly followed to plan, conduct, and report a SMS. Considering that the tasks to be carried out, presented in this protocol, adhere to these guidelines, it is believed that the execution phase (conducting the SMS) will be repeatable, and the validity threats will have been reduced as much as possible, at fully acceptable levels.

Regarding the objective of the SMS, we believe that, by following this protocol, it will be successfully achieved. The results we hope to find with this study will help out in a better selection of estimation methods in agile methodologies, in addition to providing solid evidence for other related studies.